\documentclass[aps,prl,reprint,superscriptaddress,intlimits]{revtex4-2}
\usepackage{bm,latexsym,mathrsfs,enumerate,amsmath,amssymb}

\usepackage{graphicx}
\usepackage[breaklinks=true,urlcolor = blue,colorlinks = true,citecolor = blue,linkcolor = blue]{hyperref}
\usepackage[utf8]{inputenc}

\begin{document}

\title{Weaving Hopfions from Emergent Monopoles in a Chiral Magnet}

\author{Shoya Kasai}
  \email{Contact author: kasai@aion.t.u-tokyo.ac.jp}
  \affiliation{Department of Applied Physics, The University of Tokyo, Hongo, Tokyo 113-8656, Japan}
\author{Kotaro Shimizu}
  \affiliation{Department of Applied Physics, The University of Tokyo, Hongo, Tokyo 113-8656, Japan}
\author{Shun Okumura}
  \affiliation{Quantum-Phase Electronics Center (QPEC), The University of Tokyo, Tokyo 113-8656, Japan}
  \affiliation{RIKEN Center for Emergent Matter Science (CEMS), Wako, Saitama 351-0198, Japan}
  \affiliation{International Institute for Sustainability with Knotted Chiral Meta Matter (WPI-SKCM$^2$), Hiroshima University, Hiroshima 739-8531, Japan}
\author{Yukitoshi Motome}
  \email{Contact author: motome@ap.t.u-tokyo.ac.jp}
  \affiliation{Department of Applied Physics, The University of Tokyo, Hongo, Tokyo 113-8656, Japan}

\begin{abstract}
Recent advances in three-dimensional magnetization imaging techniques have opened new avenues for exploring topological spin textures beyond domain walls and skyrmions. Among them, magnetic hopfions are particularly promising, as their knotted topology is expected to give rise to unconventional dynamics and responses; however, their controlled creation remains challenging. Here we propose a simple mechanism for generating hopfions from magnetic torons, three-dimensional textures hosting an emergent monopole-antimonopole pair. Using Landau-Lifshitz-Gilbert simulations, we show that an electric current drives the annihilation of this pair, converting a toron into a hopfion. The initial toron length determines the number of generated hopfions, while the current direction selects the sign of the Hopf invariant. We further find that the threshold current depends sensitively on material parameters, indicating a close connection to skyrmion dynamics. Our results establish an experimentally accessible route to hopfion creation and reveal a pathway from monopole defects to knotted topological textures.
\end{abstract}
\maketitle
% \tableofcontents

%%%%%%%%%%%%%%%%%%%%%%%%%%%%%%%%%%%%%%%%%%%%%%%%%%%%%%%%%%%%%%%%%%%%%%%%%%%%%%%%%%%%%%%%%%%%%%%%%%%%%%%%%%%%%%%%%%%%%%%%%%%%%%%%%%%%

Topology has become an indispensable tool for exploring functionality in condensed-matter systems, opening avenues toward efficient devices beyond the limits of conventional electronics. Topological physics has been broadly applied both in momentum space, exemplified by the quantum Hall effect and topological insulators~\cite{Klitzing1980,Thouless1982,Kane2005,Bernevig2006,Konig2007}, and in real space, including liquid crystals and Bose-Einstein condensates~\cite{Fukuda2011,Ackerman2017_nmat,Ackerman2017_PRX,Tai2019,Wu2022,Kawaguchi2008,Leslie2009,Hall2016}. Magnetic media provide another prominent platform, where topological spin textures such as domain walls and skyrmions are promising building blocks for next-generation spintronics~\cite{Barnes2006,Parkin2008,Jonietz2010,Fert2013,Zhang2015,Koshibae2015}. Recent advances in three-dimensional (3D) magnetization imaging have further expanded this field toward genuinely 3D topological textures~\cite{Park2014,Donnelly2017,Seki2022,Wolf2022,Yu2022,Henderson2023,Yu2024}.

\begin{figure}[b!]
  \centering
  \includegraphics[width=\hsize]{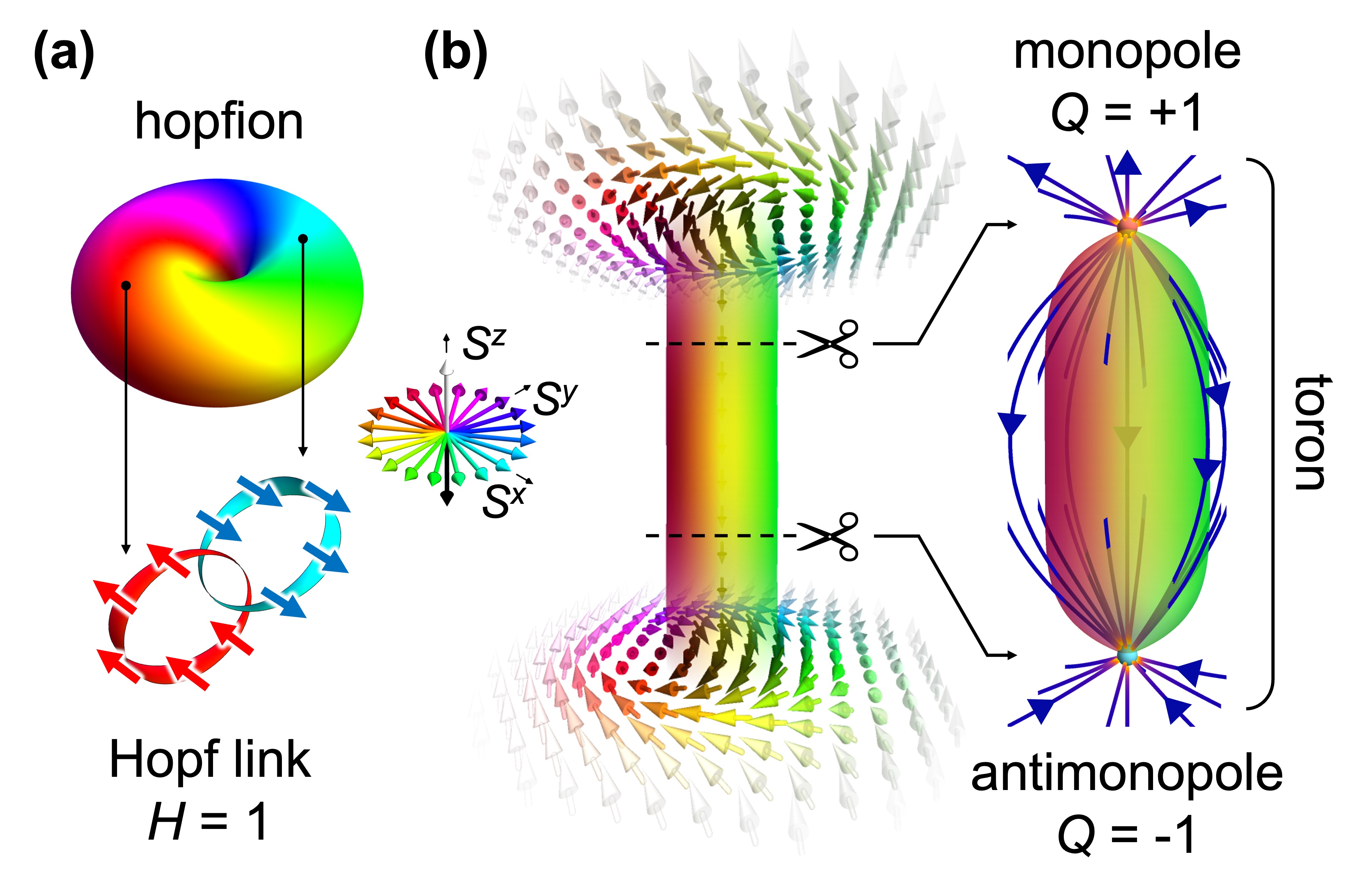}
  \caption{(a) A hopfion (top) and its two preimages (bottom). The preimages form a Hopf link that characterizes the hopfion topology with the Hopf number $H = 1$. (b) A skyrmion tube (left) and a toron (right). The toron is obtained by cutting the skyrmion tube at two points, where an emergent monopole and antimonopole are located. As illustrated by the blue streamlines, these defects serve as a source and a sink of the emergent magnetic field, which carry monopole charges $Q=+1$ and $-1$, respectively. The color code placed between (a) and (b) is used throughout this paper to represent spin configurations.}
  \label{schematic}
\end{figure}

\begin{figure*}[t!]
  \centering
  \includegraphics[width=\hsize]{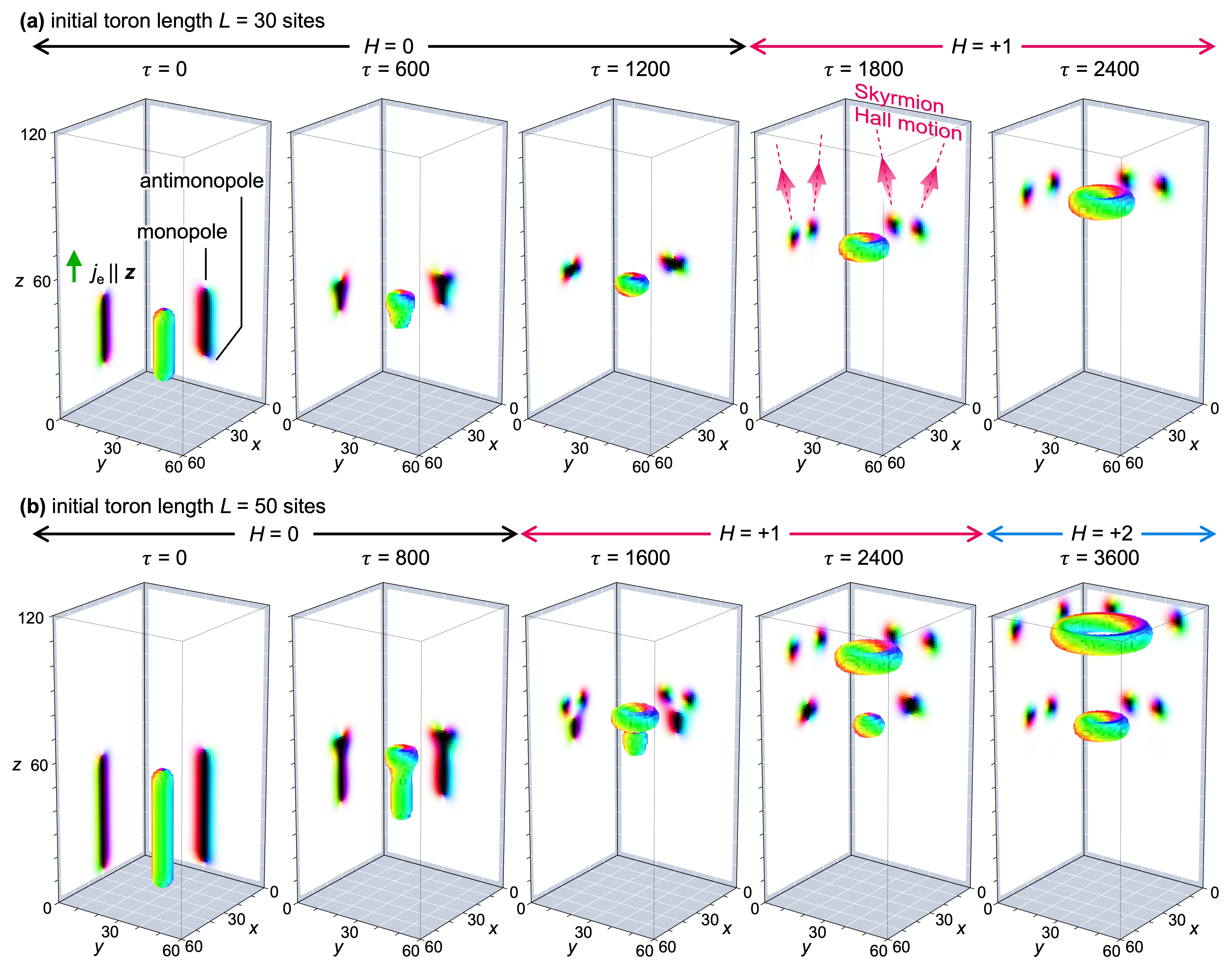}
  \caption{Current-driven dynamics of torons with initial lengths: $L = 30$ in (a) and $L = 50$ in (b), calculated for $j_{\rm e} = 0.4$ and $\beta = 0.05$. Each panel shows the 3D spin structure and its 2D projections. In the 3D views, only regions satisfying $S^z \lesssim 0$ are displayed, while empty regions correspond to $S^z \simeq 1$. The two boundary faces show the 2D configurations at $x=30$ and $y=30$. In the rightmost panel of (b), the entire structure is shifted by $30$ lattice sites along the $z$ direction for clarity.}
  \label{transition}
\end{figure*}

Among such textures, hopfions have attracted considerable attention for their unique knot topology. As shown in Fig.~\ref{schematic}(a), a hopfion exhibits a toroidal isosurface of the $z$ component of spins $S^z$, whose topology is characterized by linked preimages of identical spin orientations. The corresponding topological invariant, known as Hopf number $H$, is given by the linking number of these preimages~\cite{Hopf1931,Whitehead1947,Vega1978,Wilczek1983}. Theoretical studies have revealed a variety of intriguing phenomena, including superstructure formation~\cite{Kasai2025,Hou2025}, quantum transport~\cite{Gobel2020,Pershoguba2021,Saji2023,Gobel2025}, and unconventional dynamics~\cite{Wang2019,Liu2020,Raftrey2021,Khodzhaev2022,Liu2022,Kasai2026B,Kasai2026open}, which are unrealizable in other magnetic states. Following the recent real-space observation of hopfions, experimental verification of these predictions has become a major challenge~\cite{Kent2021,Yu2023,Zheng2023,Li2026,Chen2026.4,Chen2026.5}. A key obstacle, however, remains: how to create hopfions in a controlled manner. Experimental realizations reported to date rely either on probabilistic nucleation or on elaborate control of external fields, while a simpler and deterministic protocol for preparing hopfions is still lacking. Theoretically, current-induced hopfion creation in a cylindrical geometry was proposed~\cite{Liu2024}, where hopfions are generated as part of a mixed texture rather than as isolated objects.

The creation of hopfions is challenging because their discrete topological invariant forbids continuous deformation from topologically trivial states. Any topological transition must therefore pass through singular configurations where the invariant becomes ill-defined, analogous to band-gap closing at Dirac points in momentum space~\cite{Su1979,Qi2006,Bernevig2006}. In the real-space spin texture of 3D magnets, analogous singularities arise as emergent Dirac monopoles and antimonopoles at which the spin length vanishes, as illustrated in Fig.~\ref{schematic}(b). These defects appear at the branches and ends of skyrmion tubes, which are 3D extensions of skyrmions along the out-of-plane direction, and are referred to as hedgehogs, Bloch points, or chiral bobbers~\cite{Milde2013,Schutte2014,Rybakov2015,Kagawa2017,Zheng2018,Birch2021,Henderson2023}. Monopole-antimonopole (MAM) pairs, also known as torons~\cite{Leonov2018,Muller2020}, have been reported to stabilize as periodic arrays in several materials~\cite{Kanazawa2011,Ishiwata2011,Kanazawa2012,Tanigaki2015,Fujishiro2019,Ishiwata2020}. These considerations naturally raise a question: Is it possible to generate hopfions by driving emergent monopole dynamics? Indeed, in nematic liquid crystals and magnetic media, it has been pointed out that MAM-pair annihilation within a toron can leave behind a hopfion~\cite{Chen2013,Liu2018,Li2022,Gao2024,Souza2025}. Dynamically inducing such pair annihilation may therefore provide a promising route to hopfion creation. Although monopole dynamics has been studied previously, hopfion generation has not been reported~\cite{Hu2021,Shimizu2025,Kuchkin2026}, presumably because hopfions were not metastable within the models considered. Indeed, hopfion stabilization requires a careful balance of magnetic frustration and sample-shape anisotropy~\cite{Vakulenko1979,Bogolubsky1988,Sutcliffe2017_frustration,Sutcliffe2018,Tai2018,Naya2022,Rybakov2022,Sallermann2023,Lobanov2023}.

In this Letter, we numerically demonstrate current-driven hopfion creation from emergent monopoles in a chiral magnet. We show that an electric current parallel to a toron induces annihilation of a MAM pair, converting the toron into a hopfion rather than a topologically trivial state. For sufficiently long torons, multiple hopfions are created from one end before the annihilation. We further clarify the threshold current required for the hopfion creation by performing the simulations over a range of physical parameters. Our findings provide a simple monopole-mediated mechanism for creating hopfions and motivate further experimental exploration of hopfion physics.

We consider a 3D chiral magnet composed of $N = 61^2 \times 121$ spins under periodic boundary conditions. The Hamiltonian in Eq.~\eqref{eq:model} includes competing exchange interactions, the Dzyaloshinskii-Moriya interaction (DMI), and Zeeman coupling. Spin dynamics is investigated by solving the Landau-Lifshitz-Gilbert (LLG) equation in Eq.~\eqref{eq:LLG} including spin-transfer torque (STT) in Eq.~\eqref{eq:STT}. Details of the model and numerics are provided in the End Matter.

The leftmost panels of Figs.~\ref{transition}(a) and \ref{transition}(b) show the initial state used in our LLG simulation, each containing a single toron with MAM separation $L$. Its central axis is placed at $(x,y)=(30,30)$. The initial toron state is obtained by energy minimization using gradient-based method with automatic differentiation, starting from an ansatz of a skyrmion tube terminated at two points (see Ref.~\cite{Kasai2025} for numerical details). The model parameters in Eq.~\eqref{eq:model} represent one choice from a parameter region in which both the toron and an $H=1$ hopfion are metastable, i.e., both correspond to local minima of the energy landscape.

First, we show that a hopfion can be created through current-driven toron dynamics. Figures~\ref{transition}(a) and \ref{transition}(b) show snapshots of the toron dynamics at several simulation times $\tau$ under an electric current $j_{\rm e}=0.4$ applied along the $z$ direction, namely, parallel to the toron. The nonadiabatic coefficient associated with the STT term in Eq.~\eqref{eq:LLG} is $\beta=0.05$. Each panel displays both the 3D texture and the two-dimensional cross section on the $x=30$ and $y=30$ planes, shown on the boundary faces on $x=0$ and $y=0$, respectively. As shown in Fig.~\ref{transition}(a) for the $L=30$ case, the electric current gradually shortens the toron over $0 \leq \tau \leq 1200$. Simultaneously, the cross-sectional textures reveal an expansion of the toron in the radial direction, most clearly visible at $\tau=600$ where one end becomes noticeably broadened. As a consequence of the competition between length shortening and radial expansion, the MAM pair annihilates; however, the toron transforms into a toroidal texture rather than a trivial state, as shown in the figure at $\tau = 1800$.

To confirm its topological nature, we compute the Hopf number $H$, whose integral form is defined by~\cite{Hopf1931,Whitehead1947,Vega1978,Wilczek1983}
\begin{align}
    H = -\int \bold{B}_{\rm em}(\bold{r}) \cdot \bold{A}_{\rm em}(\bold{r})~d\bold{r} \in \mathbb{Z},
    \label{eq:hopfnum}
\end{align}
where $\bold{B}_{\rm em}(\bold{r})$ is the emergent magnetic field and $\bold{A}_\mathrm{em}(\bold{r})$ represents the corresponding vector potential satisfying $\nabla\times\bold{A}_\mathrm{em}(\bold{r}) = \bold{B}_\mathrm{em}(\bold{r})$ at position $\bold{r} = (x, y, z)$. The $\alpha$ component of $\bold{B}_\mathrm{em}$ is given by
\begin{align}
    B_{\rm em}^\alpha(\bold{r}) =\frac{1}{8\pi} \varepsilon_{\alpha\beta\gamma}\bold{S}(\bold{r}) \cdot \left\{\partial_{\beta} \bold{S}(\bold{r}) \times \partial_{\gamma} \bold{S}(\bold{r})
    \right\},
    \label{eq:Bem}
\end{align}
with the Levi-Civita symbol $\varepsilon_{\alpha\beta\gamma}$ and classical spin $\bold{S}(\bold{r})$ in continuous space. Details of the calculation on a discrete lattice are given in Ref.~\cite{Liu2018,Kasai2025}. For the texture at $\tau = 1800$, we find $H\simeq 1$, confirming that the resulting state is indeed a hopfion.

After its creation, the hopfion propagates along the current direction while expanding its toroidal radius. This expansion is due to the skyrmion Hall motions of the skyrmions with different topological charges appearing in cross section of the hopfion~\cite{Liu2020}. Remarkably, the sign of $H$ for the created hopfion can be controlled by the current direction: the positive current $j_{\rm e} > 0$ used in Fig.~\ref{transition}(a) produces an $H=1$ hopfion, whereas a negative current $j_{\rm e} < 0$ generates an $H=-1$ hopfion, as detailed in the End Matter.

\begin{figure}[t!]
  \centering
  \includegraphics[width=\hsize]{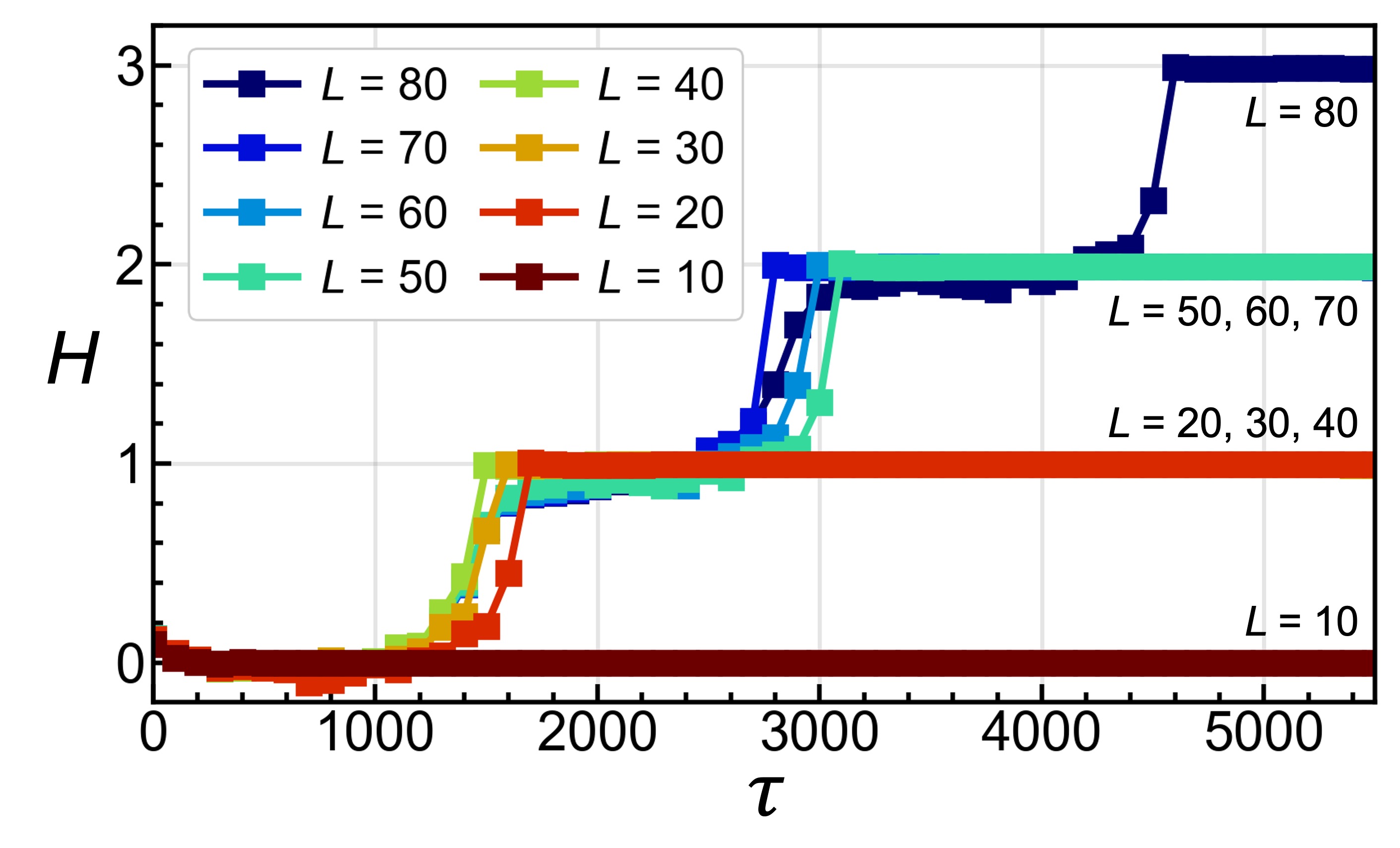}
  \caption{Time evolution of the Hopf number in Eq.~\eqref{eq:hopfnum} for different initial toron length $L$. We set $j_{\rm e} = 0.4$ and $\beta = 0.05$.}
  \label{length_hopf}
\end{figure}

Next, we investigate the dependence of the current-driven toron dynamics on the initial toron length. Figure~\ref{transition}(b) shows snapshots for $L=50$. Similar to the structure at $\tau=600$ in Fig.~\ref{transition}(a), the toron end undergoes pronounced inflation, visible at $\tau=800$ in Fig.~\ref{transition}(b). The subsequent evolution, however, is markedly different. Instead of immediately annihilating the MAM pair, the inflated region pinches off from the toron's end and forms an isolated hopfion, while the MAM pair remains in the system, as shown at $\tau=1600$. Upon further evolution, the remaining toron follows a process similar to that observed in Fig.~\ref{transition}(a), eventually resulting in two $H=1$ hopfions state at $\tau = 3600$.

Figure~\ref{length_hopf} shows the time evolution of the Hopf number for different initial toron lengths. For $L=10$, shown in brown, the MAM pair annihilates before the toron's end can appreciably expand, relaxing the system into a topologically trivial ferromagnetic state. In contrast, for longer torons, one or more hopfions are generated prior to pair annihilation, as reflected in the stepwise evolution of the Hopf number. This behavior originates from the presence of the MAM pair, which makes the Hopf number ill-defined; as a result, the Hopf number is no longer constrained to integer values and can evolve continuously. Once the MAM pair annihilates, it becomes well-defined and locked to a stable integer value~\cite{note_H_evolution}.

\begin{figure}[t!]
  \centering
  \includegraphics[width=\hsize]{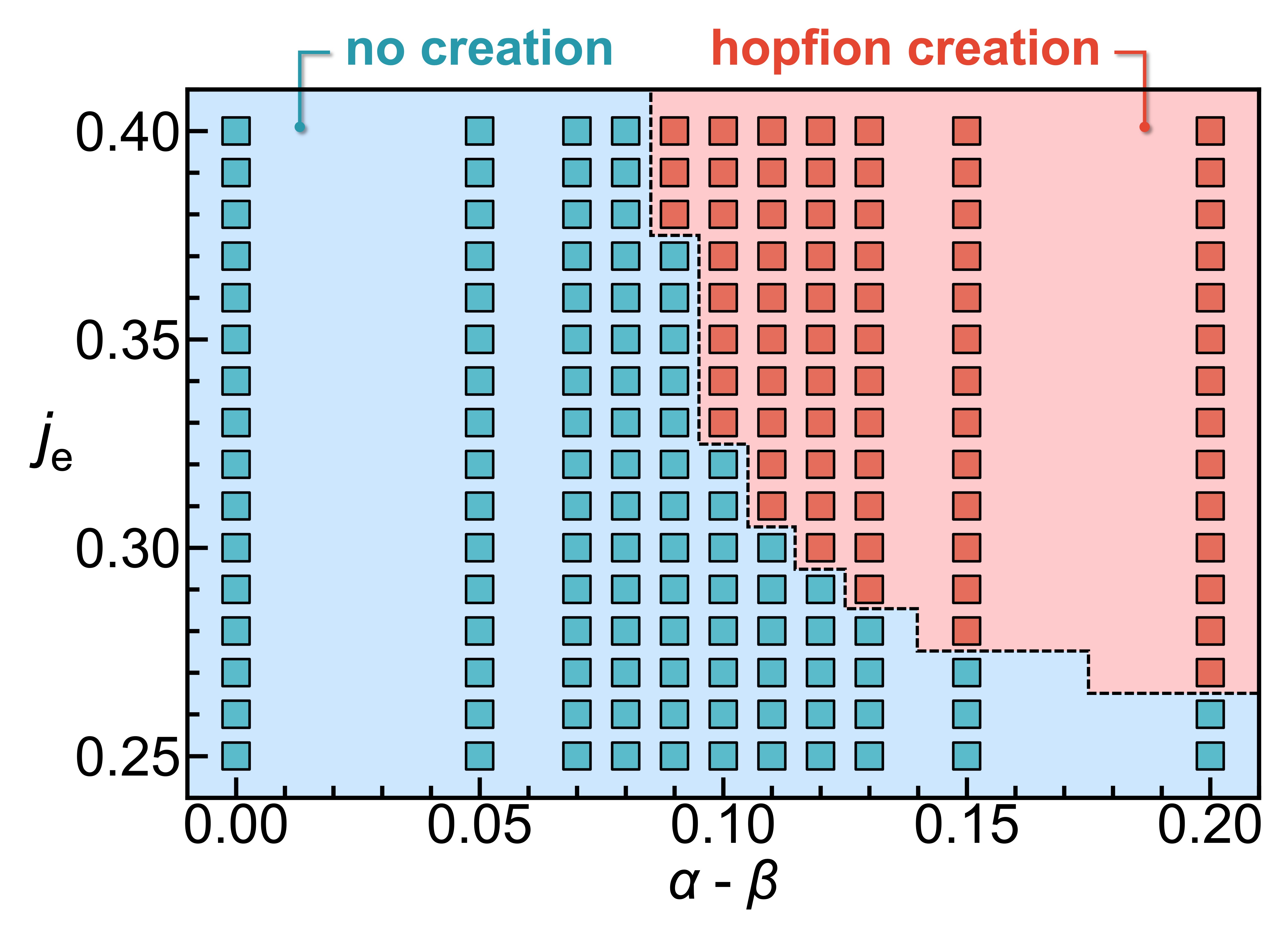}
  \caption{Final-state diagram at $\tau = 7000$ for fixed damping $\alpha = 0.2$. Squares represent simulated parameter sets, while the dashed line denotes the phase boundary. In the red-shaded region, hopfion creation occurs within the simulation time range, whereas in the blue-shaded region the MAM annihilates and the system relaxes to a topologically trivial state. The final states are distinguished by calculating the Hopf number in Eq.~\eqref{eq:hopfnum}.}
  \label{diagram}
\end{figure}

Suppose that the inflation of a toron’s end serve as a precursor to the skyrmion Hall motion observed after hopfion creation. Then, knowledge of skyrmion dynamics may provide insight into the mechanism underlying the onset of hopfion formation. In particular, because the skyrmion Hall angle is governed by $\alpha-\beta$~\cite{Nagaosa2013}, a larger $\alpha-\beta$ is expected to enhance the inflation of the toron's end and, in turn, facilitate hopfion creation. To test this, we perform LLG simulations for a toron with $L=30$, varying $j_{\rm e}$ and $\beta$ at fixed Gilbert damping $\alpha = 0.2$. In the red region of Fig.~\ref{diagram}, an $H=1$ hopfion is generated by $\tau=7000$, whereas no hopfion creation occurs in the blue region. The black dashed line represents the threshold current for hopfion creation, showing that larger $\alpha - \beta$ allows hopfions to be generated at lower currents. We also confirm that the toron-to-hopfion transition occurs at earlier times for larger $\alpha - \beta$ (not shown). In blue regions with small $\alpha - \beta$ or low current densities $j_{\rm e}$, the expansion of the toron's end is suppressed, and the spin state instead relaxes to a topologically trivial ferromagnetic state through MAM pair annihilation. These results demonstrate that the cross-sectional skyrmion Hall motion plays a key role in enabling the stable generation of hopfions.

Finally, we discuss the rationale behind this toron-to-hopfion conversion protocol. Our focus on current-driven dynamics was inspired by a previous study of two-dimensional textures, where an electric current applied to a pinned spin texture was shown to generate a skyrmion-antiskyrmion pair~\cite{Karin2017}. Since a cross section of a hopfion contains a skyrmion-antiskyrmion pair, this result suggests that applying a current to an axis-symmetric 3D spin texture may provide a route to hopfion generation. Such a precursor must contain singularities where the Hopf number in Eq.~\eqref{eq:hopfnum} is ill-defined, which led us to focus on torons. In this context, a chiral bobber with a single monopole or antimonopole provides a closely related candidate, as it can be viewed as a toron with a large separation between its MAM pair, similar to that investigated in Fig.~\ref{length_hopf}. This strategy resembles the generation of 3D hydrodynamic vortex rings, such as dolphin bubble rings and volcanic smoke rings, suggesting a potential analogy between fluid dynamics and spin-texture dynamics.

In summary, we discovered the current-driven creation dynamics of hopfions from torons. For short torons, the current rapidly induces annihilation of the MAM pair, driving the system into a topologically trivial state. In contrast, for torons with intermediate length, the current causes not only shortening of the toron but also inflation of one end; the competition between these processes results in the production of a hopfion. For longer torons, the end expansion reaches a critical stage before MAM annihilation occurs, leading to detachment of the inflated region as an isolated hopfion. Consequently, multiple hopfions can be generated from a single long toron. Because the inflation of the toron's end can be regarded as a precursor to the expansion of the toroidal radius of the generated hopfion, the critical current required for the hopfion creation depends sensitively on the physical parameters as shown in Fig.~\ref{diagram}.

Our findings suggest several intriguing directions for future research. While the present study focused on a single toron, it would be worthwhile to investigate the dynamics of toron crystal phases found in cubic magnets~\cite{Kanazawa2011,Tanigaki2015,Fujishiro2019,Ishiwata2020} toward the realization of hopfion crystal phases~\cite{Kasai2025,Hou2025}. Since the essence lies in the presence of singularities, alternative routes to hopfion creation beyond the current-driven mechanism also deserve further investigation. With an eye toward spintronic applications of hopfions, it would also be interesting to explore the functionalities enabled by the combination of the hopfion creation with the fusion~\cite{Ward2000,Hietarinta2012,Rybakov2022}, splitting~\cite{Kasai2026B,Kasai2026open}, and deletion dynamics~\cite{Liu2020}. Furthermore, extending similar creation protocols to other systems hosting singular defects is a fascinating topic. Our discovery not only accelerates experimental studies of hopfion physics but also stimulates interdisciplinary interest in the connection between monopole physics and knotted textures.

%%%%%%%%%%%%%%%%%%%%%%%%%%%%%%%%%%%%%%%%%%%%%%%%%%%%%%%%%%%%%%%%%%%%%%%%%%%%%%%%%%%%%%%%%%%%%%%%%%%%%%%%%%%%%%%%%%%%%%%%%%%%%%%%%%%%

% \section*{\label{acknowledge}Acknowledgments}
We thank Kaito Kobayashi and Shuichi Murakami for fruitful discussions. This work was supported by the JSPS KAKENHI (No.~JP22K13998, No.~JP23K25816, No.~JP25H01247, and No.~JP26H00634) and JST PRESTO (No.~JPMJPR2595). S. K. was supported by the Program for Leading Graduate Schools (MERIT-WINGS) and JST SPRING, Grant Number JPMJSP2108. The computation in this work has been done using the facilities of the Supercomputer Center, the Institute for Solid State Physics, The University of Tokyo.

%%%%%%%%%%%%%%%%%%%%%%%%%%%%%%%%%%%%%%%%%%%%%%%%%%%%%%%%%%%%%%%%%%%%%%%%%%%%%%%%%%%%%%%%%%%%%%%%%%%%%%%%%%%%%%%%%%%%%%%%%%%%%%%%%%%%

\bibliography{bibliography}

\appendix

\newpage
\onecolumngrid
\vspace{\columnsep}
\begin{center}
\rule[3pt]{0.4\textwidth}{0.4pt}
\textbf{\large{ \ End Matter \ }}
\rule[3pt]{0.4\textwidth}{0.4pt}
\end{center}
\vspace{\columnsep}
\twocolumngrid

{\it Model and Method}
---We consider a 3D chiral magnet with competing spin interactions on a simple cubic lattice. The Hamiltonian is given by
\begin{align}
  \mathcal{H} = &-\sum_{\alpha = 1}^4 \sum_{\langle i,j \rangle_\alpha} J_{\alpha}~\bold{S}_i \cdot \bold{S}_j + \sum_{\langle i,j \rangle_1} \bold{D}_{i,j} \cdot (\bold{S}_i \times \bold{S}_j) \nonumber \\
  &- B \sum_{i}S_i^z,
\label{eq:model}
\end{align}
where $\bold{S}_i = (S_i^x,S_i^y,S_i^z)$ represents the classical spin with $|\bold{S}_i|=1$ at site $i$. The first term represents the exchange interactions up to the fourth-neighbor pairs; the summation with respect to $\langle i,j \rangle_{\alpha}$ runs over $\alpha$th-neighbor pairs. We take $(J_1,J_2,J_3,J_4) = (1,-0.166,0,-0.083)$, similar to the parameters used in previous studies on hopfions~\cite{Bogolubsky1988,Liu2020,Rybakov2022}. The magnetic frustration is introduced to circumvent the limitation imposed by Derrick's theorem~\cite{Derrick1964}. The second term denotes the DMI between nearest-neighbor pairs. The DM vector is taken along the bond direction $\bold{D}_{i,i+\mu} = D\hat{\bm \mu}$, where $\hat{\bm \mu}$ is the unit vector along the $\mu = x, y, z$ direction. The third term describes the Zeeman coupling to the magnetic field. In this study, we set $D=0.01$ and $B=0.006$. For these parameter values, the ground state is ferromagnetic along the $z$ direction. We set the lattice constant to unity and impose periodic boundary conditions.

To investigate real-space and real-time spin dynamics, we solve the LLG equation given in dimensionless form,
\begin{align}
  \frac{d \bold{S}_i}{d\tau} &= \frac{1}{1 + \alpha^2} (\bold{S}_i \times \bold{H}_i^{\rm eff} + \alpha \bold{S}_i \times (\bold{S}_i \times \bold{H}_i^{\rm eff}) + \tilde{j_{\rm e}} \bold{T}_i),
  \label{eq:LLG}
\end{align}
where $\tau$ and $\alpha$ represent the dimensionless time and the Gilbert damping constant, respectively. The effective field is defined as $\bold{H}_i^{\rm eff} = \frac{\partial \mathcal{H}}{\partial \bold{S}_i}$. The spin polarized electric current is given as $\tilde{j_{\rm e}} = p j_{\rm e}/2$, where $p$ is the spin polarization of the current and $j_{\rm e}$ is the current amplitude. The current-induced STT is given by~\cite{Zhang2004}
\begin{align}
  \bold{T}_i = (\beta& - \alpha)\bold{S}_i \times \partial_z \bold{S}_i \nonumber \\
  &+ \alpha \beta \bold{S}_i \times (\bold{S}_i \times \partial_z \bold{S}_i) - \partial_z \bold{S}_i,
\label{eq:STT}
\end{align}
where $\beta$ is the nonadiabatic coefficient. On the discrete lattice, the spatial derivative is evaluated using the central-difference approximation, i.e., $\partial_z \bold{S}_i = (\bold{S}_{i + z} - \bold{S}_{i - z})/2$. In the continuum limit, the second term in Eq.~\eqref{eq:STT} reduces to $-\alpha \beta (\bold{j}_{\rm e} \cdot \bm{\nabla}) \bold{S}_i$ because $|\bold{S}_i| = 1$ and $\bold{S}_{i} \cdot (\bold{j}_{\rm e} \cdot \bm{\nabla}) \bold{S}_i = 0$. We set $\alpha = 0.2$ and $p = 0.2$ throughout this work. We numerically integrate Eq.~\eqref{eq:LLG} using the fourth-order Runge-Kutta method with time step $\Delta \tau = 0.1$. The units of $\tau$ and $j_{\rm e}$ correspond to $\hbar/J \simeq 0.66$~ps and $eJ/\hbar a^2 \simeq 10^{12}$~A/m$^2$, respectively, for the Dirac constant $\hbar = 6.58 \times 10^{-16}~\rm{eV \cdot s}$, an energy unit $J = 1$~meV, the elementary charge $e = 1.60 \times 10^{-19}$~C, and a lattice constant $a = 0.5$~nm~\cite{Iwasaki2013}.

\begin{figure*}[t!]
  \centering
  \includegraphics[width=\hsize]{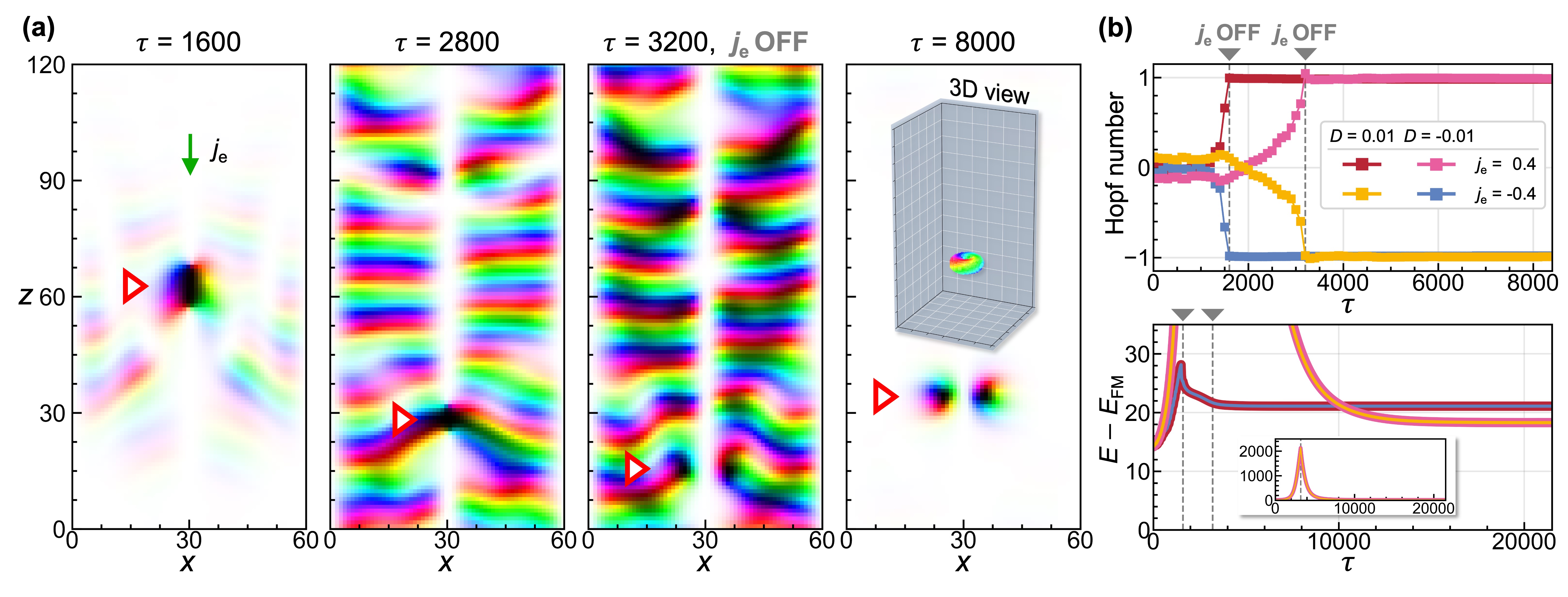}
  \caption{(a) The snapshots of the cross section at $y=30$ plane during the toron dynamics driven by a pulse current with $j_{\rm e}=-0.4$. The current is switched off at $\tau=3200$. All simulation conditions except the current direction are the same as those in Fig.~\ref{transition}(a), therefore the initial state is omitted here. The red triangle in each panel marks the position of the tracked spin texture. (b) The upper panel shows the time evolution of the Hopf number for the four combinations of current direction and the sign of the DMI. For the cases shown in red and blue, the current is switched off at $\tau=1600$ to compare the energy of the created hopfions. The lower panel shows the corresponding time evolution of the energy, with the inset displaying the magenta and yellow curve over a wider vertical range.}
  \label{endmatter}
\end{figure*}

{\it Dynamics under a current in the negative $z$ direction}
---Figure~\ref{endmatter}(a) displays the time evolution of the spin configuration on the $y=30$ plane under an electric current applied along the negative $z$ direction with $j_{\rm e}=-0.4$. All simulation parameters, except for the current direction, are identical to those used in Fig.~\ref{transition}(a). The initial state is also the same as that in Fig.~\ref{transition}(a) and is therefore omitted from Fig.~\ref{endmatter}(a).

Similar to the transient dynamics shown in Fig.~\ref{transition}(a), toron shortening occurs at $\tau=1600$ in Fig.~\ref{endmatter}(a). However, no discernible inflation of the toron's end is observed; instead, a helical modulation begins to emerge in the background, with its chirality selected by the sign of $D$. As shown in the panel at $\tau=2800$, this modulation gradually develops while preserving the MAM pair. At $\tau=3200$, the MAM pair annihilates, and the toron cross section appears to split into two black parts resembling that of a hopfion. We then examine the relaxation dynamics from this hopfionlike state by switching off the current at $\tau = 3200$ (otherwise, the system evolves into a disordered state). Since the ground state of the present spin model is ferromagnetic, the helical modulation gradually fades away, leaving only the aforementioned texture by $\tau=8000$.

To investigate its topological nature, we show the time evolution of the Hopf number $H$ in the upper panel of Fig.~\ref{endmatter}(b) for the $j_{\rm e}=-0.4$ case (yellow), together with the relaxation dynamics just after hopfion creation for $j_{\rm e}=0.4$ (red) as a reference. This confirms that the remnant at $\tau = 8000$ is the $H = -1$ hopfion. Thus, a hopfion with a desired sign of $H$ can be created by selecting the current direction.

We also present the simulation results for the opposite sign of $D$ in the same panel. In this case, for $j_{\rm e}=0.4$ (magenta), a helical instability with chirality opposite to that in the yellow case is induced, generating an $H=1$ hopfion. In contrast, under $j_{\rm e}=-0.4$ (blue), expansion of the toron's end leads to an $H=-1$ hopfion creation.

In the lower panel of Fig.~\ref{endmatter}(b), we present the time evolution of the total energy $E$ for these four dynamics, where $E$ is measured from the ferromagnetic-state energy $E_{\rm FM}/N=-1.761$. After a sufficiently long time, the system reaches a steady state in each case, revealing the energy difference between the $H=\pm1$ hopfions originating from the DM interaction in the $z$ direction. In the hopfion-creation processes shown in magenta and yellow, a helical instability develops and causes a significant increase in energy during the transient regime. Since the chirality of this excited helical state is favored by the DMI, the created hopfion after relaxation has a lower energy.

\end{document}